\documentclass[prb,aps,amsmath,amssymb,reprint,showpacs,floatfix,superscriptaddress]{revtex4-1}
\usepackage{graphicx}
\usepackage{dcolumn}
\usepackage{bm}
\usepackage{mathtools}

\begin{document}


\title{Effects of tilting the magnetic field in 1D Majorana nanowires} 



\author{Javier Osca}
\email{javier@ifisc.uib-csic.es}
\affiliation{Institut de F\'{\i}sica Interdisciplin\`aria i de Sistemes Complexos
IFISC (CSIC-UIB), E-07122 Palma de Mallorca, Spain}
\author{Daniel Ruiz}
\affiliation{Institut de F\'{\i}sica Interdisciplin\`aria i de Sistemes Complexos
IFISC (CSIC-UIB), E-07122 Palma de Mallorca, Spain}
\author{Lloren\c{c} Serra}
\affiliation{Institut de F\'{\i}sica Interdisciplin\`aria i de Sistemes Complexos
IFISC (CSIC-UIB), E-07122 Palma de Mallorca, Spain}
\affiliation{Departament de F\'{\i}sica,
Universitat de les Illes Balears, E-07122 Palma de Mallorca, Spain}

\date{February 25, 2014}

\begin{abstract}
We investigate the effects that a tilting of the magnetic field 
from the parallel direction has
on the states of a 1D Majorana nanowire. Particularly, we focus on the conditions for the existence of Majorana zero modes,
uncovering an analytical relation (the sine rule) between the field orientation 
relative to the wire, its magnitude and the superconducting parameter of the material.
The study is then extended to junctions of nanowires,
treated as 
magnetically inhomogeneous straight nanowires composed of two homogeneous arms. It is shown that their spectrum can be explained in terms of the spectra of two independent arms. Finally, we investigate how the localization of the Majorana mode is 
transferred from the magnetic interface at the corner of the junction
to the end of the nanowire when increasing the arm length.
\end{abstract}


\maketitle 
\section{Introduction}
\label{intro}

In 2003 Kitaev pointed out the usefulness of topological states for 
quantum computing operations.\cite{Kitaev} Essentially, topological states are quantum states with a hidden internal symmetry.\cite{Affleck} They are usually localized close to the system edges or interfaces and their nonlocal nature gives them a certain degree of immunity against local sources of noise. A subset of this kind of states called Majorana edge states is attracting much interest in condensed matter 
physics.\cite{Wilczeck,Qi,Alicea,Leijnse,Beenakker,StanescuREV,Franz,Fu,Akhmerov,Tanaka,Law} Majorana states are effectively chargeless zero-energy states that behave as localized non abelian anyons. 
It is theorized that nontrivial phases arise from their mutual interchange, caused by their nonlocal properties.\cite{Nayak,Pachos} Furthermore, these states have the property of being their own anti-states, giving rise to statistical behavior that is neither fermionic nor bosonic. Instead, the creation of two Majorana quasiparticle excitations in the same state returns the system to its equilibrium state. This kind of quasiparticles inherits its name from Ettore Majorana who theorized the existence of fundamental particles with similar statistical properties.\cite{Majorana}

Majorana states have been theoretically predicted in many different systems and some of them have been realized experimentally. In particular, evidences of their formation at the ends of
semiconductor quantum wires inside a magnetic field with strong spin-orbit interaction and in close proximity to a superconductor
have been seen in Refs.\ \onlinecite{Mourik,Deng,Rokhinson,Das,Finck}.
Superconductivity breaks the charge symmetry creating quasiparticle states without a defined charge that are a mixture of electron and hole excitations. On the other hand, the spin-orbit Rashba effect is caused by an electric field perpendicular to the propagation direction that breaks the inversion symmetry of the system while the external magnetic field breaks the spin rotation symmetry of the nanowire. 
The combined action of both effects makes the resulting state effectively spinless and, including superconductivity, also effectively chargeless and energyless.\cite{Lutchyn,Oreg,Stanescu,Flensberg, Potter,Potter2,Ganga,Egger,Zazunov,Prada,Klino1,Klino2,Lim,Lim2,Lim3,Serra,Osca}  

This work addresses the physics of 1D nanowires with varying relative orientations between the external magnetic field
and the nanowire (see Fig.\ \ref{F1}). 
This physics is of relevance, e.g.,  for the exchange of Majoranas on networks 
of 1D wires, where it has been suggested that Majoranas can be braided by 
manipulating the wire shapes and orientations.\cite{Alicea2,Tewari,Halperin}  
The Hamiltonian of 
the system is expressed in the continuum and the 
analysis is performed using two complementary approaches: the complex band structure of the homogeneous wire
and the numerical diagonalization for finite systems. The complex band structure allows a 
precise characterization of the 
parameter regions of the semi-infinite wire where Majoranas, if present, are not
distorted by finite size effects. On the contrary
numerical diagonalizations of finite systems, even though reflecting the same underlying physics, 
yield smoothened transitions between different physical regions of parameter space.

For the semi-infinite system we uncover an analytical law limiting the existence of Majorana modes below
critical values of the angles between the magnetic field and the nanowire. This law, referred to in this article as the sine rule, is shown to be approximately valid in finite systems too. We find a 
correspondence of the finite system spectrum with its infinite wire counterpart, explaining this way its distinctive features and regimes in simplest terms. 

The results for the homogeneous nanowire are subsequently used to explain the 
spectrum of a
junction of two nanowires with arbitrary angle. The junction is modeled as a non homogeneous straight nanowire with two regions characterized by different magnetic field orientations 
(see Fig.\ \ref{F1}b).
While the magnetic field remains parallel to the nanowire in one arm, we study the spectrum variation when changing the magnetic field angles in the other. Similarities between the homogeneous and inhomogeneous nanowire spectra allow us to explain many of the features of the latter in terms of those of the former. Finally, we investigate the dependence with the distance of the magnetic interface (the corner of the junction) to the end of the nanowire, finding a transfer phenomenon where the Majoranas change localization from the interface for a short arm
to the nanowire end as the arm length is increased.

This work is organized as follows. In Sec.\ \ref{sec:1} the physical model is introduced and 
 Sec.\ \ref{sec:2} presents the above mentioned sine rule.
In Sec.\ \ref{sec:3} we discuss the spectrum of excited states of a homogeneous nanowire while in Sec.\ 
\ref{sec:4} we address an inhomogeneous system representing a nanowire junction. 
We study changes in the spectrum due to the tilting (\ref{subsec:4a}) 
and stretching (\ref{subsec:4b}) of one of the junction arms. 
Finally, the conclusions of the work can be found in Sec.\ \ref{conclusions}.

\section{Physical model}
\label{sec:1}

We assume a one dimensional model of a semiconductor nanowire as a
low energy representation of a higher dimensional wire with lateral extension, 
when only the first transverse mode is active.  
The system is described by a Hamiltonian of the Bogoliubov-deGennes kind,
\begin{eqnarray}
	\mathcal{H}_{\it BdG} &=& 
	\left( \frac{p_{x}^{2}}{2m} + V(x) -\mu \right)\tau_z \nonumber\\
	&+& \!\! \Delta_B\, \left(\sin\theta \cos\phi\, \sigma_x 
	+ \sin\theta \sin\phi\, \sigma_y +\cos\theta\, \sigma_z \right ) \nonumber\\
	&+& \Delta_s\, \tau_{x}+\frac{\alpha}{h}\, p_x \sigma_{y}\tau_z\;,
	\label{E1}
\end{eqnarray}
where the different terms are, in left to right order, kinetic, electric
and chemical potential, Zeeman, superconducting and Rashba spin-orbit terms. 
The Pauli operators for spin are represented by $\sigma_{x,y,z}$ while those for isospin are given by $\tau_{x,y,z}$. 
Superconductivity is modeled as an s-wave superconductive term that couples different states of charge.

The superconductor term in Eq.\ (\ref{E1}) is an effective mean field approximation to a more complicated phonon-assisted attractive interaction between electrons. This interaction leads to the formation of Cooper pairs with break-up energy $\Delta_s$. Experimentally, superconductivity can be achieved by close proximity between the semiconductor nanowire and a metal superconductor. The semiconductor wire becomes superconducting  when its width is smaller than the coherence length of the Cooper pairs. On the other hand, the Rashba spin orbit term arises from the self-interaction between an electron (or hole) spin with its own motion. This self interaction is due to the presence of a transverse electric field that is perceived as an effective magnetic field in the rest frame of the quasiparticle. This electric field can be induced externally but, usually, is a by-product of an internal asymmetry of the nanostructure. In the Hamiltonian Eq.\ (\ref{E1}) we have taken $\hat{x}$ as the orientation of the 1D nanowire while an effective spin orbit magnetic field $\vec{B}_{so}$ pointing along ${\hat{y}}$ may be defined 
due to the coupling of the Rashba term with the $y$ component of the spin.

\begin{figure}
\centering
\resizebox{0.4\textwidth}{!}{%
	\includegraphics{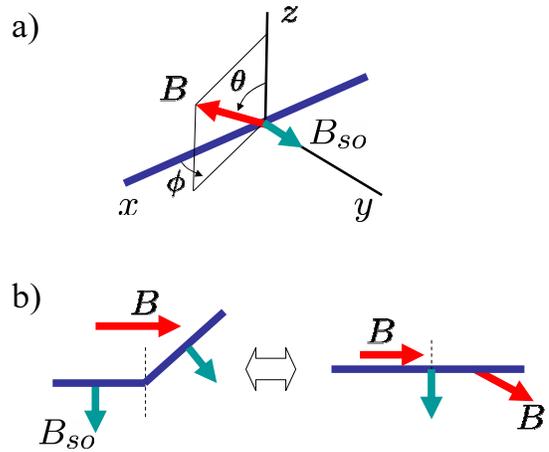}%
}
\caption{Sketches of the physical systems considered in this work. 
a) Straight nanowire on the $x$ axis
in a homogeneous magnetic field characterized by spherical angles $\theta$ and $\phi$. 
b) Junction of two nanowires in a uniform magnetic field (left) represented as a 
straight nanowire with a magnetic inhomogeneity (right).
}
\label{F1}
\end{figure}

We consider the nanowire in an external magnetic field,
giving spin splittings through
the Zeeman term in Eq.\ (\ref{E1}). In this paper we assume the magnetic field in arbitrary direction, including the possibility of being inhomogeneous in space for some setups.
The direction of the magnetic field is parametrized by the spherical polar and azimuthal angles $\theta$ and $\phi$. 
These two angles are constant for a homogeneous wire (Fig.\ \ref{F1}a)
and they change smoothly from one to the other arm in a 
nanowire junction (Fig.\ \ref{F1}b).


Summarizing, superconductor, Rashba spin-orbit and Zeeman effects are parametrized in Eq.\ (\ref{E1}) by $\Delta_s$, $\alpha$ and $\Delta_B$, respectively. These parameters are taken constant because the nanowire is considered to be made of an homogeneous material. The only inhomogeneity allowed in certain cases is a change in the magnetic field direction at a single magnetic interface between two homogeneous regions.

Along this work the Hamiltonian of Eq.\ (\ref{E1}) is solved for homogeneous 
parameters in the
infinite, semi-infinite and finite wires, as well as for the inhomogeneous finite case, 
using different approaches. When a direct diagonalization of the Hamiltonian for a finite system is performed, soft potential edges and magnetic interface are used. The shape of the potential edges is modeled as Fermi-like functions centered on those edges. High potential is imposed outside the nanowire while low potential (usually zero) is assumed inside. When a magnetic interface is present a smooth variation
in the field angles is modeled in the same way. Specifically, those
smooth functions read
\begin{eqnarray}
	\label{E2}
	V(x) & = & V_0\left[1+{\cal F}(x; x_L, s_v)-{\cal F}(x; x_R, s_v)\right]\;,\\
	\theta(x)&=& \theta_L+(\theta_R-\theta_L)\left[1-{\cal F}(x; x_m, s_m)\right]\; ,\\
	\phi(x)&=& \phi_L+(\phi_R-\phi_L)\left[1-{\cal F}(x; x_m, s_m)\right]\; ,
\end{eqnarray}
for the potential and the field polar and azimuthal angles, respectively. The Fermi function ${\cal F}$ is defined as
\begin{equation}
{\cal F}(x; x_0, s)=\frac{1}{1+e^{(x-x_0)/s}}\; .
\end{equation}
In Eq.\ (\ref{E2})
$V_0$ is the value of the potential outside the nanowire while 
$\theta_{L/R}$ and $\phi_{L/R}$ 
are the field angles at left and right of the magnetic interface. 
The potential left and right edges are centered on $x_L$ and $x_R$
and the magnetic interface is centered on $x_m$. Their softness is controlled by the parameters $s_v$ and $s_m$, where zero softness means a steep interface and a high value implies a smooth one.

The numerical results of this work are presented in special units obtained by taking 
 $\hbar$, $m$ and  the Rashba spin-orbit interaction $\alpha$ as reference values. 
That is,
our length and energy units are 
\begin{eqnarray}
	L_{\it so}&=&\frac{\hbar^2}{\alpha m}\,,\\
	E_{\it so}&=&\frac{\alpha^2 m}{\hbar^2}\,.
	\label{E3}
\end{eqnarray}

\section{A sine rule}
\label{sec:2}

\begin{figure}
\centering
\resizebox{0.5\textwidth}{!}{%
	\includegraphics{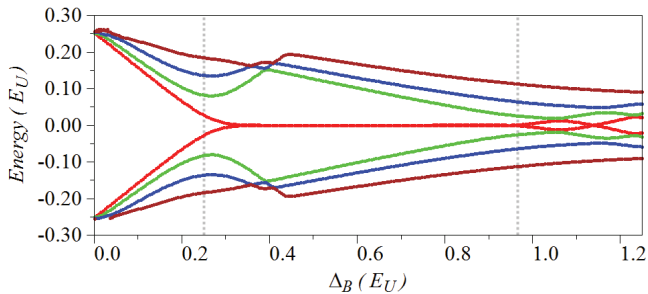}
}
\caption{ 
Spectrum of a finite length nanowire with $L=50L_{so}$ as a function of the external magnetic field magnitude $\Delta_B$. 
Other nanowire parameters are $\Delta_s=0.25E_{so}$ and $\mu=0$. The magnetic field angles are $\theta=90^{\circ}$ and $\phi=15^{\circ}$. 
Only the eight states lying closer to zero energy are displayed. 
Note that a zero energy Majorana mode is created at around $\Delta_B=0.3E_{so}$ and destroyed for values of $\Delta_B$ near one unit.
The vertical lines (dots) indicate the onset and destruction of the Majorana mode as 
predicted by  Eqs.\ (\ref{E4})  and (\ref{E7b}), respectively. 
}
\label{F2}
\end{figure}

Let us consider a nanowire in a uniform magnetic field with 
$\theta=90^\circ$ and an arbitrary $\phi$ (see Fig.\ \ref{F1}a). 
A direct diagonalization of Eq.\ (\ref{E1}) for a finite length of the wire
and $\phi=15^\circ$  yields the spectrum 
depicted in Fig.\ \ref{F2} as a function of 
the magnetic field intensity. 
A main feature of this figure is the existence of a Majorana mode, lying 
very near zero energy,  
but only for a particular range of values of the magnetic field. For the parameters 
of the figure the Majorana mode is created 
around $\Delta_B=0.3 E_{so}$ and destroyed in a rather abrupt way
around $\Delta_B= E_{so}$.

It is well known that Majorana wave functions decay to zero towards 
the nanowire interior. We can therefore analyze the creation and destruction of Majoranas in
the semi-infinite system and use those results to understand the physics of Majoranas in
a finite system. In this approach we eliminate from the analysis the finite size effects caused by the overlapping of the Majorana wave functions at both ends of a finite nanowire. 
Although it is obvious that for long enough wires the size effect becomes negligible, disentangling finite size behavior from intrinsic Majorana physics using calculations of only finite systems
is much less obvious. 

\begin{figure}
\centering
\resizebox{0.4\textwidth}{!}{%
	\includegraphics{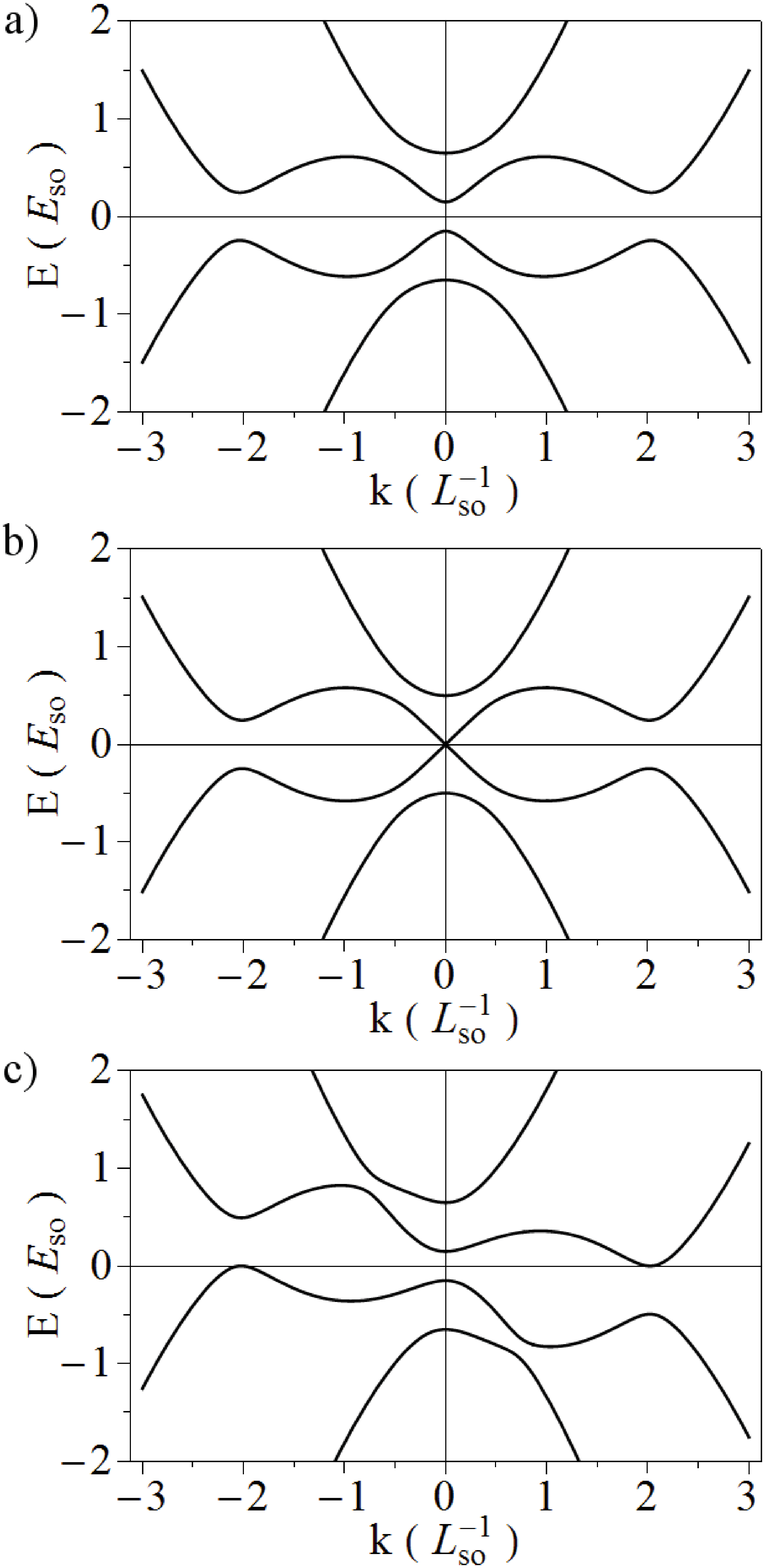}
}
\caption{Band structure of the infinite homogeneous nanowire with $\Delta_s=0.25E_{so}$
and $\mu=0$ for 
a) parallel field $(\theta,\phi)=(90^{\circ}, 0^\circ)$ with $\Delta_B=0.4E_{so}$,
b) the same as a) but on the phase transition point $\Delta_B=0.25E_{so}$,
c) tilted field $(\theta,\phi)=(90^{\circ},38.68^\circ)$ with $\Delta_B=0.4E_{so}$.
}
\label{F3}
\end{figure}

Majorana mode creation has been understood as a phase transition of the lowest excited state, signaled by the closing and reopening of a gap in the infinite nanowire band spectrum,\cite{Oreg} as shown in Figs. \ref{F3}a and \ref{F3}b. The phase transition follows in this case a well known law, requiring high-enough fields for Majoranas to exist, 
\begin{equation}
	\Delta_B\geq\sqrt{\Delta_s^2+\mu^2}\;.
	\label{E4}
\end{equation}
Notice that, as mentioned, for the equality in Eq.\ (\ref{E4}) a gap closes
for $k=0$ in Fig.\ \ref{F3}b.
 
In Ref.\ \onlinecite{Serra} Eq.\ (\ref{E4}) was derived, in an alternative way, from the analysis of the 
complex-$k$ solutions
compatible with the boundary condition of a semi-infinite nanowire in a
parallel field. 
This approach relies on the property that the complex band structure (allowing an imaginary
part in $k$)
of the homogeneous wire contains all the information about all possible eigenstates
of any piecewise homogeneous wire.
In general, an eigenstate of the infinite homogeneous wire with a given arbitrary $k$
can be expressed as
\begin{equation}
	\Psi^{(k)}(x,\eta_\sigma,\eta_\tau)=\sum_{s_\sigma,s_\tau} \Psi_{s_\sigma s_\tau}^{(k)}
	\, e^{ikx}\, \chi_{s_\sigma}(\eta_\sigma)\chi_{s_\tau}(\eta_\tau)\;,
	\label{E5}
\end{equation}
where $\Psi_{s_\sigma s_\tau}^{(k)}$ are state amplitudes and the quantum numbers are $s_\sigma=\pm$ and $s_\tau=\pm$.

The sharp semi-infinite wire with $x>0$ is obviously piecewise homogeneous, implying that 
the Majorana solution allowed by the existence of an edge at $x=0$ must be a linear superposition of the homogeneous nanowire eigenstates of complex wave number with ${\rm Im}(k)>0$,  otherwise it could not be a localized state. The resulting restriction is
\begin{equation}
	\sum_{k,\, {\rm Im}(k)>0}C_k \Psi_{s_\sigma s_\tau}^{(k)}=0\;,
	\label{E5b}
\end{equation}
where the $C_k$'s are complex numbers characterizing the superposition of state amplitudes. 
The allowed wave numbers are calculated solving the determinant 
\begin{equation}
	{\rm det}\left\{H_{s_\sigma s_\tau,s'_\sigma s'_\tau}(k)-E\, \openone \right\}=0\;,
	\label{E6}
\end{equation}
for $E=0$. In fact, the allowed $k$'s can be calculated for any energy but we are interested in particular in those at zero energy corresponding to Majorana solutions. 

\begin{figure}
\centering
\resizebox{0.4\textwidth}{!}{%
	\includegraphics{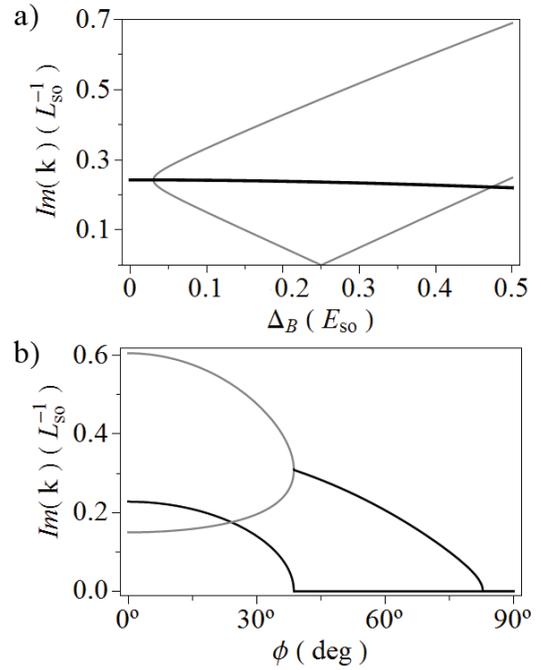}
}
\caption{Imaginary parts of the wave numbers (only positive ones) in an
infinite homogeneous nanowire with 
$\Delta_s=0.25E_{so}$ and $\mu=0$ as function of 
a) the value of the longitudinally oriented magnetic field,   
b) the azimuthal angle $\phi$ of a magnetic field with  $\Delta_B=0.4E_{so}$ and polar angle $\theta=90^{\circ}$.
Gray color is used for non degenerate modes while black is indicating degeneracy  with
two or more modes actually having
the same ${\rm Im}(k)$.
}
\label{F4}
\end{figure}

The wave number dependence on magnetic field is depicted in Fig.\ \ref{F4}a
for a selected case. For a fixed energy $E$ there are always eight possible 
wave numbers, but only those with ${\rm Im}(k)>0$ are displayed in Fig.\ \ref{F4}a. 
In this representation the closing of the $k=0$ gap in Fig.\ \ref {F3}b 
corresponds to a node of ${\rm Im}(k)$ in Fig.\ \ref{F4}a. 
In order to be able to hold a Majorana
a semi-infinite nanowire has to fulfil two simultaneous requirements. 
First, the nanowire must have four complex wave numbers with ${\rm Im}(k)>0$ 
allowed at zero energy; and second, a solution different from zero (nontrivial) must be possible for the $C_k$'s in Eq.\ (\ref{E5b}). 
That is, interpreting the state amplitudes $\Psi^{(k)}_{s_\sigma s_\tau}$
as a $4\times 4$ matrix where the four $k$'s correspond for instance  to rows and 
the four spin-isospin values $\{++,+-,-+,--\}$ to columns, the condition 
for a nontrivial solution is  
\begin{equation}
	{\rm det}\{ \Psi_{s_\sigma s_\tau}^{(k)} \}=0\;.
	\label{E7a}
\end{equation}
In a parallel field this condition is fulfilled 
only above a critical value of the magnetic field
$\Delta^{(c)}_B=\sqrt{\Delta^2+\mu^2}$,
but not under this quantity, thus leading to Eq.\ (\ref{E4}). 
Further details on the methodology can be found in Ref.\ \onlinecite{Serra}. 
Here we want to use this approach to determine whether a similar
condition on the field orientation, with critical values of the angles, 
exists or not.

Figure \ref{F4}b shows the evolution of the wave numbers 
when increasing $\phi$ while maintaining $\theta=90^{\circ}$, i.e.,
maintaining the magnetic field in the plane formed by the nanowire direction and the effective spin orbit magnetic field direction $\vec{B}_{so}$. 
This means that for $\phi=0^{\circ}$ the magnetic field is aligned with the nanowire, while for $\phi=90^{\circ}$ it is completely perpendicular to it and parallel to $\vec{B}_{so}$. 
In Fig.\ \ref{F4}b care has been taken to choose a value of $\Delta_B$
that fulfills the Majorana condition for the parallel $\phi=0$
orientation Eq.\ (\ref{E4}).
We can see that for $\phi=40.1^{\circ}$ two of the complex wave numbers 
become real, thus destroying the Majorana mode for azimuthal angles
above this value. 

\begin{figure}
\centering
\resizebox{0.5\textwidth}{!}{%
	\includegraphics{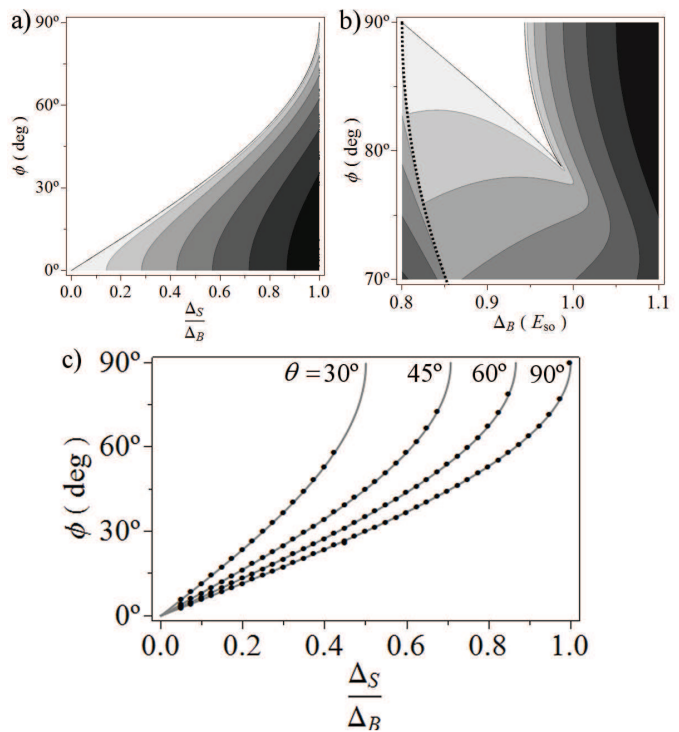}
}
\caption{ a) Contour plot of ${\rm Im}(k)$ for the external branch of the nanowire
propagating bands. The horizontal axis contains the ratio
${\Delta_s}/{\Delta_B}$ and the vertical one the azimuthal angle $\phi$. 
The polar angle is fixed to $\theta=90^{\circ}$ and $\mu=0$. b)  Contour plot of ${\rm Im}(k)$ for the internal branch of the nanowire
propagating bands. The horizontal axis shows $\Delta_B$ and the vertical one the azimuthal angle $\phi$. The polar angle is fixed to $\theta=90^{\circ}$, $\Delta_s=0.8 E_{so}$ and $\mu=0$. c) Plot of the azimuthal critical angle where ${\rm Im}(k)$ vanishes in the upper left panel as function of  ${\Delta_s}/{\Delta_B}$ (points) checked against the sine rule prediction Eq.\ (\ref{E7b}). Besides the $\theta=90^\circ$ case of the upper left panel, the figure
also contains the comparison for other values of $\theta$.
The value of the chemical potential can be taken arbitrarily since it 
is irrelevant for this comparison.
}
\label{F5}
\end{figure}

The physical behavior implied by Fig.\ \ref{F4}b is a sudden 
loss of the Majorana mode as the tilting angle $\phi$
exceeds a critical value, due to the system no longer having 
the required four evanescent modes with ${\rm Im}(k)>0$.
The evanescent modes are lost because of the closing of the gap 
between states of opposite wave numbers ($k\approx \pm2L_{\it so}^{-1}$ for the particular case shown in Fig.\ \ref{F3}c).
We characterize next the dependence of the critical angle 
on $\Delta_B$ and $\Delta_s$.
In Fig.\ \ref{F5}a we can see a contour plot of ${\rm Im}(k)$ as a function 
of $\phi$ and the ratio $\Delta_s/\Delta_B$ for an external branch wave number,\cite{Klino2} corresponding to the lower black line of Fig.\ \ref{F4}b.
The values where ${\rm Im}(k)$ vanish separate the plot into two regions, 
the lower one where the Majorana is allowed and the upper (white) where no Majorana can exist. 
Although Eq.\ (\ref{E6}) can be solved analytically, the angles where  
${\rm Im}\big(k(\phi,\Delta_s/\Delta_B)\big)$ vanishes can be obtained only numerically 
because $\phi$ appears as argument of sine and cosine functions and no isolation 
is possible. 
As a consequence, the values of $\phi$ where the wave number first reaches zero have been found numerically and are plotted in Fig.\ \ref{F5}c against the test function $\arcsin(\Delta_s/\Delta_B)$. 
The perfect coincidence between the two results within computer precision demonstrates that 
a Majorana can not exist for angles such that  $\sin\phi>\Delta_s/\Delta_B$, provided $\theta=90^{\circ}$.

Figure \ref{F5}b shows a contour plot of ${\rm Im}(k)$ for an internal branch wave number,\cite{Klino2} 
corresponding to the upper mode in Fig.\ \ref{F4}b.
In this plot 
the $\phi$ roots of ${\rm Im}(k)$ lie inside an upper and lower bounded region
around $0.95 E_{\rm so}<{\Delta_B}< E_{\rm so}$. 
In fact, two of the wave numbers become real in the white region of the contour plot. Note that this region lies in the non Majorana sector, above the transition discussed
in panel a) which is now 
signaled by the dotted line.
Theoretically the existence of this region determines two different fermionic regimes. One where a fermion mode at zero energy is constructed of plane waves with two complex and two real wave numbers and another one made of a full set of real wave numbers. Since we assume bound states in order to extrapolate the results to finite systems, these cases have no relevance to us. Nevertheless the underlying causes for the existence of this region will be relevant in the study of the excited states of the finite nanowire. This will be further developed in Sec.\ \ref{sec:3}.

Repeating the analysis for different polar angles $\theta$, 
as shown in Fig.\ \ref{F5}c, we conclude that the angular restriction 
for the existence of  Majoranas  is
\begin{equation}
	\Delta_B \sin\theta \sin\phi < \Delta_s\; .
	\label{E7b}
\end{equation}
In other words, the projection of the magnetic field 
energy parameter 
into the spin orbit effective magnetic field $\vec{B}_{so}$ 
needs to be smaller than the superconductor gap energy 
in order to have Majoranas in a semi-infinite wire.
We refer to this condition as the sine rule. 
Notice that Eq.\ (\ref{E7b}) is not a generalization of Eq.\ (\ref{E4}), but an additional law. Both Eq.\ (\ref{E4}) and Eq.\ (\ref{E7b})
have to be simultaneously met for the existence of a Majorana mode 
in a semi-infinite wire.

In general, the sine rule Eq.\ (\ref{E7b}) yields an extra bound to be considered
when identifying regions of Majoranas in parameter space.
For instance, assuming fixed angles $(\theta,\phi)$ and varying $\Delta_B$ 
there is a lower bound on $\Delta_B$ from Eq.\ (\ref{E4})
and an upper bound from the sine rule.
Analogously, if for a fixed $\Delta_B$
the Majorana is allowed 
by Eq.\ (\ref{E4}) at $\phi=0^{\circ}$ and
and we increase $\phi$
the sine rule yields an upper bound on $\phi$.
Therefore, as explained, both equations must be met simultaneously 
to obtain a Majorana mode.
Furthermore, after some parameter testing we have determined that the sine rule
is not affected by the value of the chemical potential $\mu$. 
This means that the 
overall dependence on $\mu$ for the existence of Majorana modes 
in the semi-infinite nanowire
is completely covered by Eq.\ (\ref{E4}). 

The disappearance of the Majorana when increasing 
$\phi$ is not a phase transition in the sense that 
no imaginary part of a mode wave number crosses zero in between
two regions with non null values.
As shown in Fig.\ \ref{F4}b
for the polar angle $\theta=90^{\circ}$, above the critical $\phi$
the value of 
${\rm Im}(k)$ remains stuck at zero value.  
The main difference between the phase transition law in Eq.\ (\ref{E4}) and the sine rule Eq.\ (\ref{E7b}) lies in the different type of gap closing for both cases. 
As shown in Fig.\ \ref{F3}b the phase transition delimited by Eq.\ ($\ref{E4}$) is caused by a gap closing and reopening on a single wave number $k=0$ (labeled as interior branches of the spectrum). In the language of semiconductor band structure physics
we may call this the closing of a direct gap.
Oppositely, the sine rule is caused by the closing of an indirect gap for $k\approx \pm k_f$ (labeled as exterior branches of the spectrum), as shown in Fig.\ \ref{F3}c for a selected case. 

\begin{figure}
\centering
\resizebox{0.4\textwidth}{!}{%
	\includegraphics{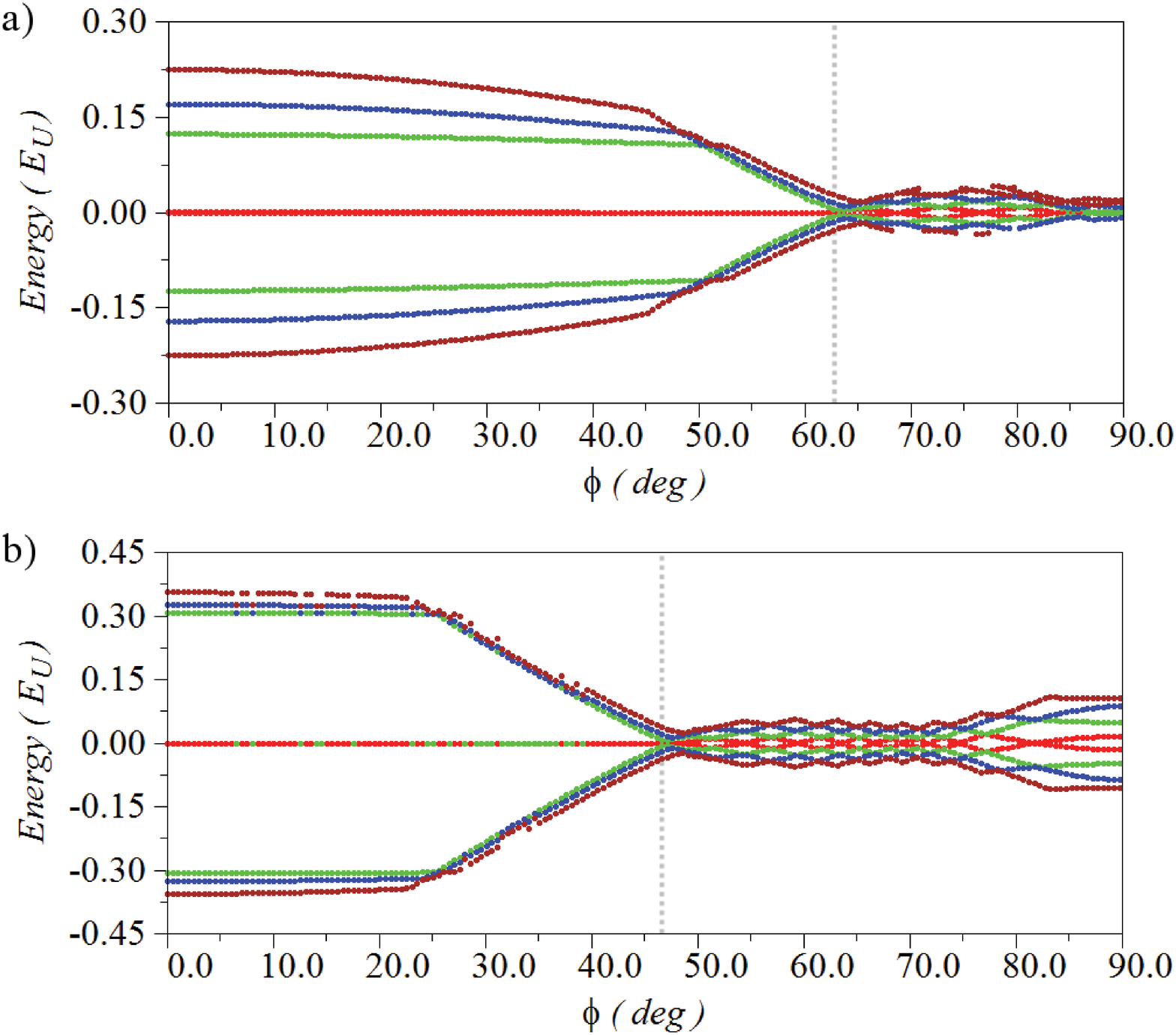}
}
\caption{ a) Spectrum of a nanowire of length $L=50 L_{\rm so}$
with $\Delta_s=0.8E_{so}$ and $\mu=0$ as a function of $\phi$ with a fixed $\theta=90^{\circ}$ and $\Delta_B=0.9E_{so}$. Note the spectrum change at the angle predicted by the sine rule
(dotted line) as well as the spectrum  
collapse for values of $\phi$ close to $90^{\circ}$. For values of $\phi$ above $90^\circ$
the spectrum is given by the mirror image of the shown values.
b) The same as a) but for a fixed magnetic field value $\Delta_B=1.1E_{so}$. Note that for values of $\phi$ close to $90^{\circ}$ now the two modes closer to zero are fermionic modes separated from each other by an energy gap.
}
\label{F7}
\end{figure}

We have checked these laws against the direct numerical diagonalization 
for a finite nanowire, finding a reasonable agreement as shown in Figs.\ \ref{F2} and \ref{F7}. 
In Fig.\ \ref{F2} the magnetic field orientation is kept fixed to a tilted orientation 
while the field magnitude is changed and in 
Fig.\ \ref{F7} the magnitude is fixed while the orientation is changed. 
The main difference between the precise laws for the
semi-infinite model and the finite system results is in the smoothness of the spectrum 
evolution around the transition points. 
While in the semi-infinite model the transition between fermionic modes to Majorana modes and 
vice versa happens at a single point in the parameter space, in the finite system we can see these transitions smoothed. This occurs due to the finite size effects, i.e., 
the little overlap of Majoranas on opposite ends of the nanowire. 
Furthermore, while Majoranas lie at exactly zero energy in the semi-infinite model, this small interaction makes the finite system Majoranas to have a finite small energy $\epsilon$.

A close inspection of the sine rule Eq.\ (\ref{E7b}) reveals that there exist critical 
values for $\theta$ and $\phi$ such that if they are not surpassed a Majorana is always allowed, independently of the value of the other angle (provided Eq.\ (\ref{E4}) is fulfilled). That is,
below the critical angle a projection into $\vec{B}_{so}$ is never high enough to break the Majorana.
In practice, if the Majorana is allowed by Eq.\ (\ref{E4}), it will survive for 
any $\phi$ provided $\theta<\theta_c$ or, alternatively, for any $\theta$ provided
$\phi<\phi_c$.
These critical angles are
\begin{equation}
	 \theta_c=\phi_c= \arcsin\left(\frac{\Delta_s}{\Delta_B}\right)\; .
	\label{E7c}
\end{equation}


\section{Excited states}
\label{sec:3}

\begin{figure}
\centering
\resizebox{0.4\textwidth}{!}{%
	\includegraphics{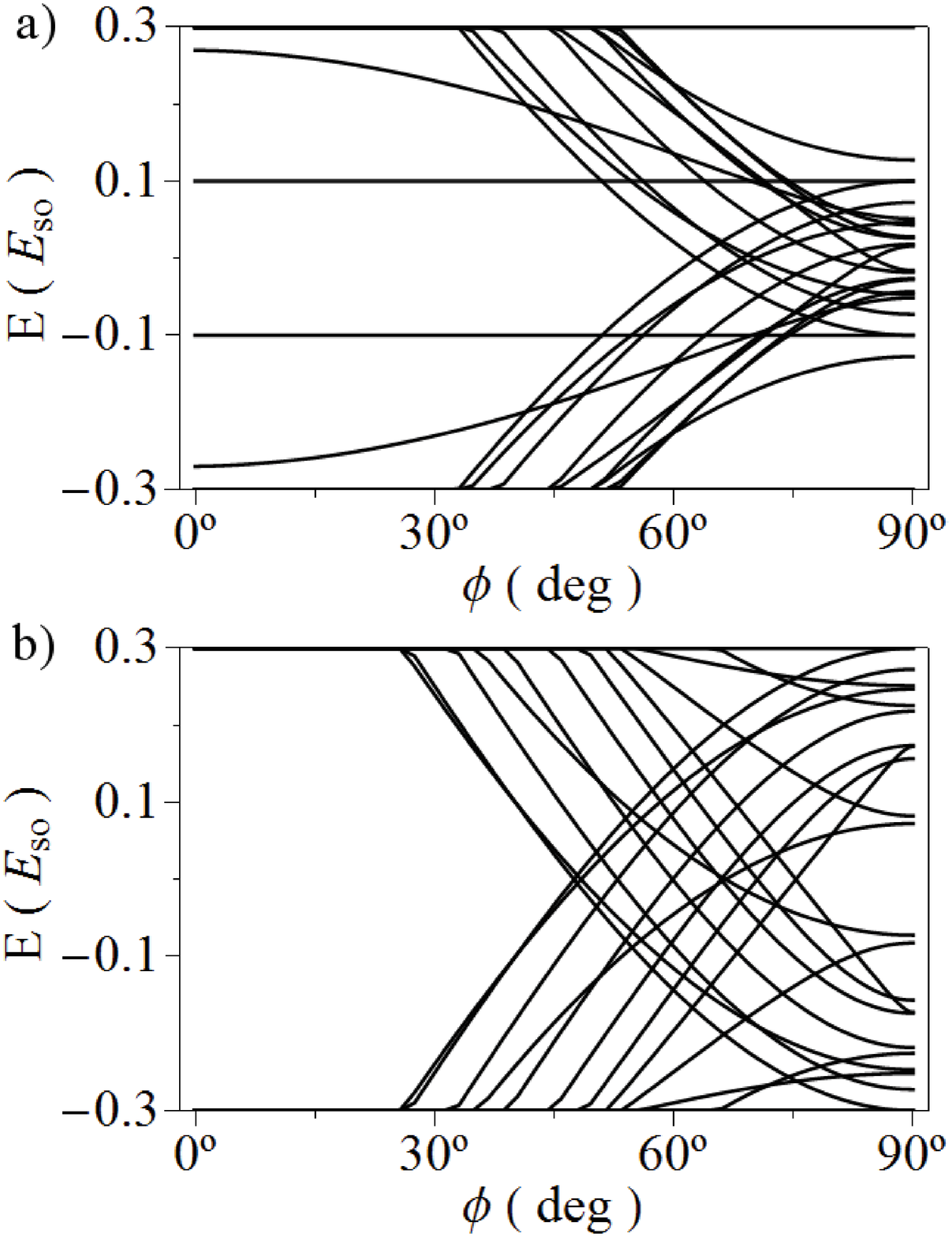}
}
\caption{Spectrum obtained
from the homogeneous nanowire band energies
at selected wave numbers, displayed as tracers as a function of $\phi$. 
Panel a) corresponds to the same parameters of Fig.\ \ref{F7}a. Note that 
qualitatively similar regions occur for increasing  angles in both figures. In particular the spectrum nearly collapses for values of $\phi$ close to $90^{\circ}$.   
Panel b) shows the same as a) but for the parameters of Fig.\ \ref{F7}b. Note that for values of $\phi$ close to $90^{\circ}$ the two lower states (closer to zero energy) are fermionic modes separated by a gap as in Fig.\ \ref{F7}b.
}
\label{F8}
\end{figure}

While in the preceding section we focussed on the physics of the Majoranas at zero energy,
comparing semi-infinite and finite nanowires, 
in this section we address the spectrum of excited states.
The main effect of the boundary conditions is to allow only a discrete set of wave numbers
instead of a continuous one.
What we have done is sketch the finite nanowire spectrum by selecting wave numbers 
at regular intervals and tracking the evolution of their energy levels 
with an increasing angle $\phi$. For these examples we maintain the polar angle $\theta=90^{\circ}$ because this is the most physically interesting configuration due to the possibility of aligning external and spin-orbit magnetic fields; 
nevertheless, analogous plots can be done for different values of $\theta$. 
The resulting spectrum, shown in Fig.\ \ref{F8}, explains the main features 
of the numerical diagonalization results of Fig.\ \ref{F7} for the same parameters. 

In principle, we could also set the boundary conditions exactly as we did in  
Eq. (\ref{E5b}), but we have found this approach impossible to follow on 
a practical level. The resulting set of equations reads
\begin{eqnarray}
	\sum_{k}C^{(L)}_k \Psi_{s_\sigma s_\tau}^{(k)}e^{-i{kL}/{2}}=0\; ,\nonumber\\
	\sum_{k}C^{(R)}_k \Psi_{s_\sigma s_\tau}^{(k)}e^{i{kL}/{2}}=0\;,
	\label{E8}
\end{eqnarray}
where $C^{(L)}_k$ and $C^{(R)}_k$ are the coefficients at the left and right nanowire ends, respectively. 
Basically, the resulting matrix from Eq.\ (\ref{E8}) is ill defined since it contains very large and very small matrix elements.

\begin{figure}
\centering
\resizebox{0.4\textwidth}{!}{%
	\includegraphics{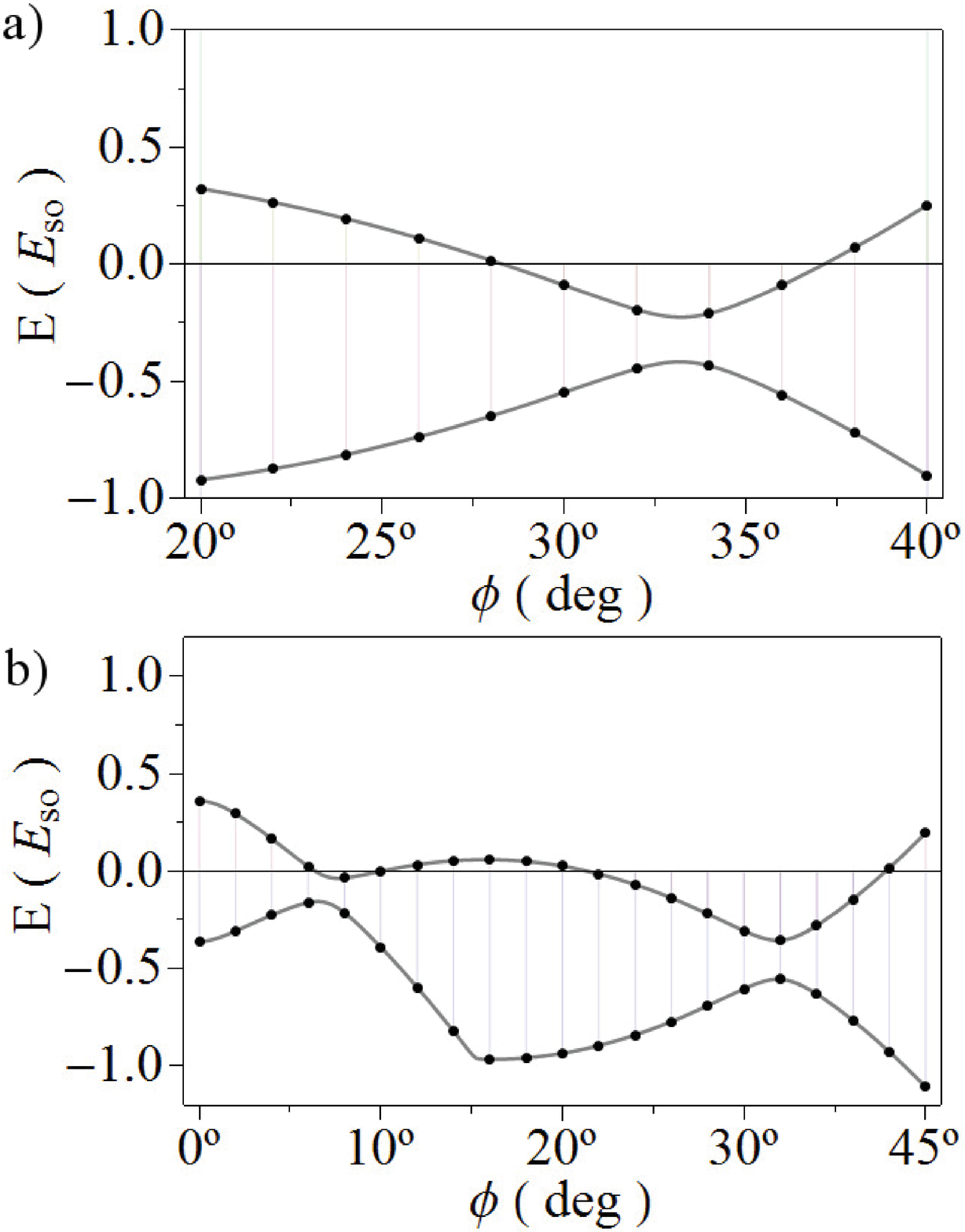}
}
\caption{ Approximate spectrum of a finite nanowire in a particular configuration of the external magnetic field. a) With only two positive real wave numbers. b) With four positive real wave numbers.
}
\label{F9}
\end{figure}

The spectra of both panels of Fig.\ \ref{F8} 
can be divided into three different regions depending on the angle $\phi$.
First, for low values of $\phi$  there is a region where a Majorana mode exists and is topologically protected. In Fig.\ \ref{F8} the Majorana is not seen since
only excited states of real wave number are shown, but we can see the corresponding gap. 
For values of $\phi$ above those determined by the sine rule the Majorana mode is destroyed and we can see a region of many level crossings.  This behavior of the spectrum is explained by the gap closing of the external branches of the conduction band noticing that in the finite model only some discrete values are allowed, as sketched in Fig. \ref{F9}. Finally, for higher angular values the region of zero crossings finishes and a third region arises with two possible behaviors. 

As shown in Figs.\ \ref{F7} and \ref{F8}
for high $\phi$ angles (near $90^{\circ}$),
depending on the parameters 
the spectrum either opens a gap 
or collapses near zero energy. The behavior depends on the way the internal branches of the band cross the zero energy value for those angles.
The internal branches of the band can cross the zero energy level for high angles in one $k>0$ point, like in Fig.\ \ref{F9}a, thus leading (jointly with the external branch crossing point) to four real and four complex wave numbers; or, alternatively, the 
interior brach can cross zero energy in more than one 
$k>0$  point, like in Fig.\ \ref{F9}b, leading to wavefunctions characterized by eight real wave numbers. In the latter case there is a wave number range where the band spectrum lies very close to zero energy, yielding this way a collapse of the finite wire spectrum.
The particular set of parameters where one or the other situation happens depends on the behavior of the  internal branches of the band structure and it is not as easily predictable as the behavior of the external branches that led to the sine rule. 
The region of values where this collapse arises coincides with the region where the allowed solutions at zero energy are made of real wave numbers only and it was already presented in Fig. \ref{F5}a for the $\mu=0$ case.


\section{Magnetic inhomogeneity models}
\label{sec:4}

In this section we explore the physics of a junction of two straight nanowires with a certain 
angle in presence of  a homogeneous magnetic field parallel to 
one of the arms, as sketched in Fig.\ \ref{F1}b.
We assume a representation of the system as a single straight 1D nanowire
containing a magnetic interface. The inhomogeneity separates two homogeneous regions with different directions 
(but the same magnitude) of 
the external magnetic field. The system is solved by numerical diagonalization, assuming a soft magnetic interface, interpreting the results by comparing with the homogeneous
nanowire discussed in  the preceding section. We focus on two specific effects, tilting and stretching of
one of the two junction arms. 

\subsection{Arm tilting}
\label{subsec:4a}

\begin{figure}
\centering
\resizebox{0.5\textwidth}{!}{%
	\includegraphics{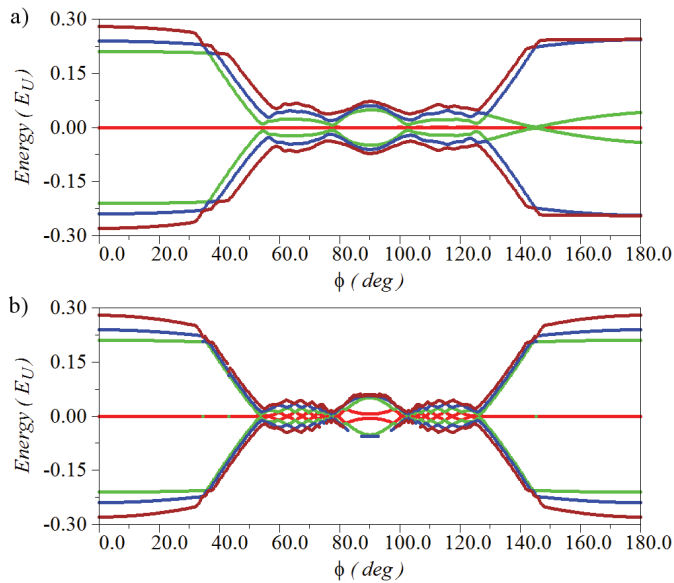}
}
\caption{ a) Nanowire spectrum for $\Delta_s=0.8E_{so}$ and $\mu=0$ with a magnetic inhomogeneity at its center as a function of the tilting angle $\phi$. On the left side of the nanowire the magnetic field 
is parallel, while on the right side its angles  are $(\theta=90^\circ,\phi)$. The magnetic field strength is constant in both sides and equal to $\Delta_B=E_{so}$. b) Spectrum of a nanowire in a homogeneous magnetic field with angles $(\theta=90^\circ,\phi)$ and with the rest of parameters as in a).
}
\label{F12}
\end{figure}

The magnetic field is aligned with the left arm and 
the spectrum of the nanowire is computed for varying orientations of the field in the the right arm 
(see Fig.\ \ref{F12}). As mentioned, this model represents 
under certain approximations a bent nanowire in a homogeneous magnetic field.  
It was shown in Ref.\ \onlinecite{Martorell} that bent nanowires can be approximated by 1D models with a potential well simulating the effect of the bending. Here we have only considered the magnetic field change of direction as the main inhomogeneity source, disregarding the electrical potential effects of  the 
bending. 

\begin{figure}
\centering
\resizebox{0.5\textwidth}{!}{%
	\includegraphics{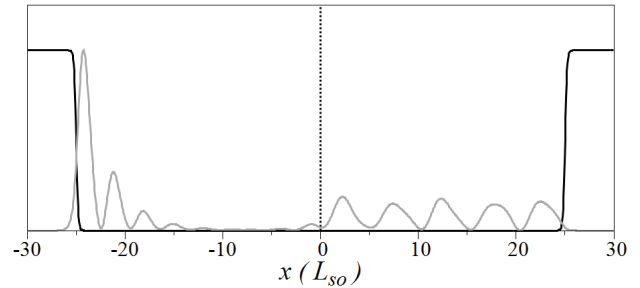}
}
\caption{Majorana density function of an inhomogeneous nanowire similar to the one described in Fig.\ \ref{F12}a. The magnetic field azimuthal angles in the two arms are 
$\phi_L=0^\circ$ and $\phi_R=90^\circ$, while all along the nanowire it is $\theta=90^\circ$.
Other parameters are $\Delta_B=0.4$, $\Delta_s=0.25$.
}
\label{F12b}
\end{figure}
\begin{figure}
\centering
\resizebox{0.5\textwidth}{!}{%
	\includegraphics{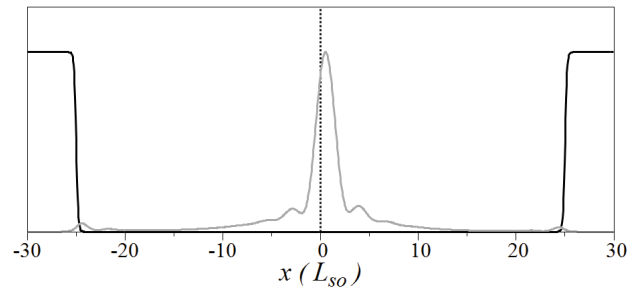}
}
\caption{Density function of the first excited state of the inhomogeneous nanowire of Fig.\ \ref{F12}a. This state becomes
localized at the magnetic inhomogeneity for an azimuthal angle $\phi=145^{\circ}$ (and polar angle $\theta=90^{\circ}$).
}
\label{F13}
\end{figure}

The spectrum of the inhomogeneous nanowire can be explained in terms of the homogeneous one  
for a tilted magnetic field. Figure \ref{F12} compares the inhomogeneous (upper) 
with the homogeneous (lower) nanowire spectrum for the same set of parameters, 
showing that both results share the same essential features. More precisely,
three $\phi$ regions can be found in both cases,  but with two main differences. First, while for the homogeneous nanowire increasing $\phi$ leads to the destruction of the Majoranas on both ends, for the inhomogeneous nanowire only the right side Majorana is destroyed. The density of the Majorana 
for  the inhomogeneous nanowire is shown in Figs.\ \ref{F12b} and \ref{F13} for selected values of the parameters.
As a consequence, the bent junction holds a Majorana mode (the one localized in the left side of the inhomogeneity) independently of the magnetic angle at the right side. 

A second difference between upper and lower panels of Fig.\ \ref{F12} is that the spectra for the inhomogeneous nanowire is not symmetric with respect to $\phi=90^{\circ}$, in contrast with the homogeneous nanowire. A zero energy crossing localized in the inhomogeneity interface arises at $\phi=145^{\circ}$ for the selected parameters in Figs.\ \ref{F12}a and \ref{F13}. The corresponding bound state originates in
the second excited state of the system and it is not Majorana in nature. Furthermore, this localized state is caused completely by the magnetic inhomogeneity and has no relationship with the localized states found in the bending region in Ref.\ \onlinecite{Martorell} because we have disregarded those effects. Although we know these states are related with the magnetic inhomogeneity, a deep understanding of their causes and the particular set of parameters leading to their enhancement or quenching is yet to be understood.

\begin{figure}[t]
\centering
\resizebox{0.5\textwidth}{!}{%
	\includegraphics{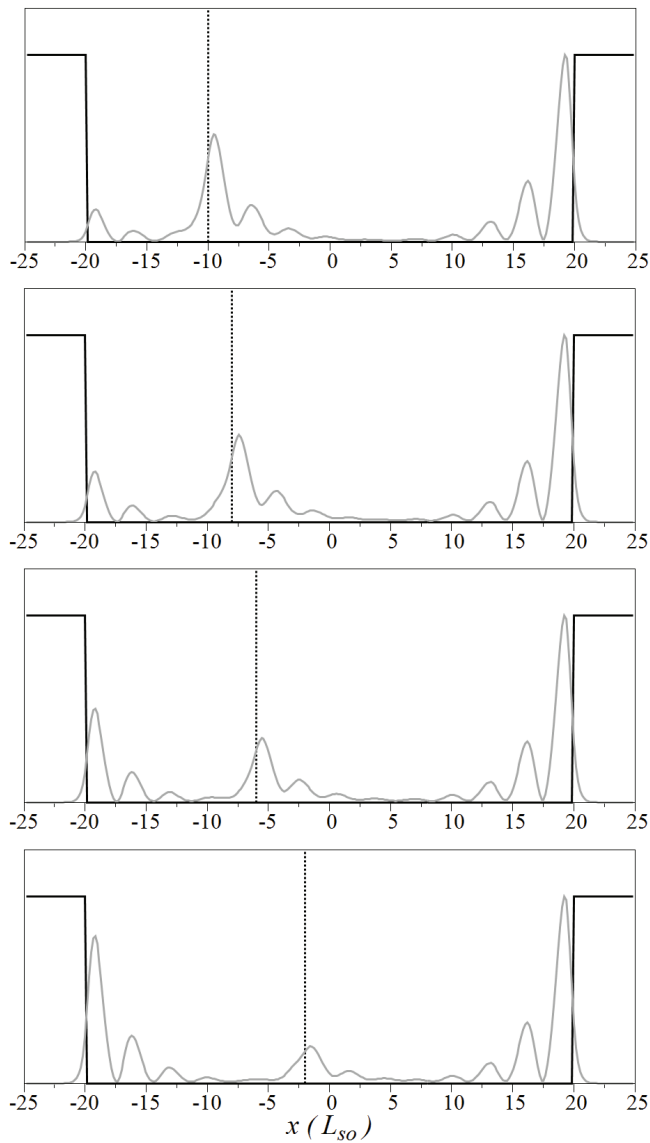}
}
\caption{Density distributions of the Majorana mode in a finite 
nanowire with a magnetic inhomogeneity.
We used $\Delta_s=0.25E_{so}$,
$\mu=0$ and a magnetic field of magnitude $\Delta_B=0.4E_{so}$ oriented parallel in the left side of the magnetic interface and 
antiparallel in
the right side. 
In each panel 
the potential well and the position of the  
magnetic interface are shown. The latter corresponds to a Fermy-type function whose
position shifts to the right following the 
sequence from upper to lower panels.
}
\label{F14}
\end{figure}

\subsection{Arm stretching}
\label{subsec:4b}

We study now the behavior of
the Majorana modes in the nanowire as a function of the inhomogeneity distance to the nanowire end. The magnetic field directions are fixed at  $(\theta=90^{\circ}, \phi=0^{\circ})$ on the left end and 
$(\theta=90^{\circ}, \phi=180^{\circ})$ on the right end of the nanowire. This is a particularly interesting configuration as it is  the only setup where both ends lie inside a longitudinal magnetic field,
apart from the homogeneous case.
 This way, all the observed effects must be caused by the inhomogeneity and its distance with respect to the left nanowire end. 

\begin{figure}[t]
\centering
\resizebox{0.5\textwidth}{!}{%
	\includegraphics{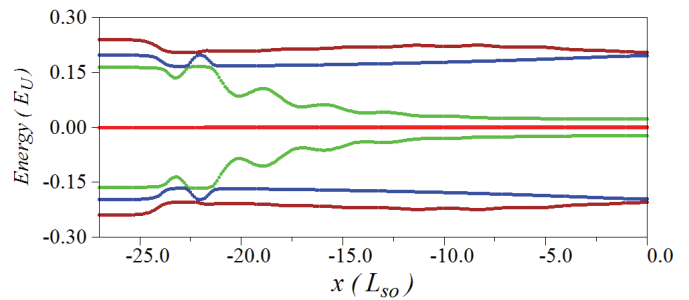}
}
\caption{ Spectrum of the nanowire described in the caption of Fig. \ref{F14} as a function the magnetic interface position.
}
\label{F15}
\end{figure}

Figure \ref{F14} shows the probability densities of the zero energy state at different positions of the magnetic interface with respect the left side of the nanowire. From upper to lower panels of Fig.\ 
\ref{F14} we may follow the evolution as the distance of the magnetic interface to the left end of the 
system is increased.
Most remarkably, for short distance the left Majorana 
is not peaked on the left end, but remains stuck on the magnetic interface (upper panels).  If the 
distance is increased, however, the Majorana is eventually transferred to the left nanowire end 
after some critical distance (lower panels). 
This transfer is seen as a smooth decrease of the density maximum at the magnetic interface accompanied by an increase at the left end. Finally, when the interface is on the middle point of the nanowire both Majoranas are  located at their corresponding ends. 
It is also worth noticing that this Majorana transfer does not imply a departure of the mode from zero energy because the Majorana on the other end of the nanowire is not affected (see Fig.\ \ref{F15}).
Additionally,  the transfer phenomenon is not caused by finite size effects since we have checked that it happens for the same characteristic distance when the right end is further displaced to the right.

\section{Conclusions}
\label{conclusions}

In this work we have studied the spectra of 1D nanowires 
for arbitrary orientation
of the magnetic field, focussing in particular on the conditions
leading to a Majorana mode. 
This study has been realized from different perspectives and methods 
in an effort to explain
the variety of observed phenomena. We have combined the complex band structure techniques of infinite homogeneous nanowires with numerical 
diagonalizations of finite systems.

We have demonstrated an additional condition, 
besides the well known topological transition law, 
that needs to be taken into account 
in order
to predict the regimes of existence of Majorana modes with tilted fields. 
We have named this additional condition the sine rule. The sine rule 
predicts an upper bound on the magnetic field at which Majoranas are to be found in a 1D wire with tilted field. 
When the topological law is fulfilled, the sine rule 
leads to 
critical values of the field angles $\theta_c$ and $\phi_c$, such that a Majorana mode is always found for any $\phi$ provided $\theta<\theta_c$
or, alternatively, for any $\theta$ provided $\phi<\phi_c$.

We have extended our analysis to nanowire junctions with an arbitrary angle, modeled as magnetically inhomogeneous nanowires, explaining most of their properties in terms of the behavior of its homogeneous parts.
We have focussed, particularly, on the role of tilting and stretching
of one of the junction arms.
We also reported the existence of a bound non Majorana state
located on the magnetic inhomogeneity. Finally, we have studied the Majorana transfer phenomenon as the distance of the magnetic inhomogeneity to the nanowire end is increased. 
Testing these predictions would require experiments of nanowires in inhomogeneous magnetic fields. Alternatively, it has been suggested in this work that a bent nanowire in a homogeneous field should display similar phenomena, while being more feasible in practice.
As an interesting continuation of this work we are presently analyzing the validity of the sine rule in higher dimensional nanowires, where 
the transverse degrees of freedom require a multimode description
of the electronic states.

\begin{acknowledgments}
This work was funded by MINECO-Spain (grant FIS2011-23526),
CAIB-Spain (Conselleria d'Educaci\'o, Cultura i Universitats) and 
FEDER. We hereby acknowledge the PhD grant provided by the University 
of the Balearic Islands.
\end{acknowledgments} 

\bibliographystyle{apsrev4-1}
\bibliography{sineCM}

\end{document}